\documentclass[aps,preprint,superscriptaddress,nofootinbib,preprintnumbers,showkeys]{revtex4-1}

\usepackage{amssymb}
\usepackage{amsmath}
\usepackage{hyperref}
\usepackage{graphicx}
\usepackage{axodraw4j}
\usepackage{pstricks}
\usepackage{color}
\usepackage{epstopdf}
\usepackage{slashed}
\usepackage{setspace}
\usepackage{caption}
\usepackage{subcaption}

\begin{document}

\preprint{HUPD1905}

\title{A New Left-Right Symmetry Model}



\author{Apriadi Salim Adam}
\email[]{apriadi.adam@gmail.com}
%\thanks{}
\affiliation{Graduate School of Science, Hiroshima University, Higashi-Hiroshima, 739-8526, Japan }%
\affiliation{Gunung Sari RT.02, RW.24 Kelurahan Ngringo, Kecamatan Jaten, Kabupaten Karanganyar, Jawa Tengah, 57731 Indonesia}%

\author{Akmal Ferdiyan}
\email[]{aferdiyan@unsoed.ac.id}
%\thanks{}
\affiliation{Department of Physics, Universitas Jenderal Soedirman, Karangwangkal Purwokerto 53123, Indonesia}

\author{Mirza Satriawan}
\email[]{mirza@ugm.ac.id (Corresponding Author)}
%\thanks{}
\affiliation{Department of Physics, Universitas Gadjah Mada, Bulaksumur Yogyakarta 55281, Indonesia}


\date{\today}

\begin{abstract}
We propose a new L-R symmetry model where the L-R symmetry transformation reverses both the L-R chirality and the local quantum number.  We add to the model a global quantum number $F$ whose value is one for fermions (minus one for anti fermion) and vanishes for bosons.   For each standard model (SM) particle, we have the corresponding L-R dual particle whose mass is very large and which should have decayed at the current low energy level.  Due to the global quantum number $F$, there is no Majorana neutrino in the model but a Dirac seesaw mechanism can still occur and the usual three active neutrinos oscillation can still be realized.  We add two leptoquarks and their L-R duals, for generating the baryon number asymmetry and for facilitating the decay of the L-R dual particles.   The decay of the L-R dual particles will produce a large entropy to the SM sector and gives a mechanism for avoiding the big bang nucleosynthesis constraint.
\end{abstract}

%\pacs{}

\maketitle


%\tableofcontents

%\newpage

\section{Introduction}
\label{intro}

The left-right symmetry (LRS) model was originally introduced based on aesthetic reason, to explain the non-existence of the right weak current \cite{patisalam2,mohapati,senjamoha}.  Even though it is no longer the center of attention in the particle physics community, there is no physical fact/observation that ruled out the LRS model as a viable standard model extension.   All fermions in the original LRS model are doublets of either SU(2)$_L$ or $SU(2)_R$ and one has to add a L-R bidoublet scalar to facilitate the mass generation of the fermions \cite{senjanovic2}.  Unfortunately this bidoublet also causes the mixing between the left and right weak gauge bosons, which is unobserved until now.  There is also a variant of the LRS model where for each SM particle one has its left (or right) chiral partner, thus doubling the SM particle content \cite{coutinho,simoes,almeida}.  In this variant model, fermions gain their mass through seesaw mechanism \cite{seesaw}, except for neutrinos that undergo a double seesaw mechanism \cite{coutinho,simoes,almeida}.  There is no requirement for bidoublet in this variant model, but the seesaw mechanism for the charged fermions is not compatible with the recent result of the Higgs scalar coupling to the SM fermions, where the coupling is proportional to the fermion mass \cite{sirunyan}.  There is also another variant of the LRS model where one doubles both the particle content and the gauge group of the SM. This model is also known as the mirror model \cite{foot3,volkas}.  There is no interaction between  SM particles and  mirror particles, except through the U(1) kinetic mixing and the scalar mixing in the scalar potentials.  This  model has a very rich hidden sector and the mirror particles can be dark matter candidates.  Since the mirror particles do not share any common quantum numbers with the SM particles, the chance to directly detect or generate any mirror particle at the accelerator is very small.

In this paper, we propose a new variant of the LRS model with a new LRS transformation and an additional global quantum number.  The new model is still invariant under the gauge group SU(3)$\otimes$SU(2)$_L\otimes$SU(2)$_R\otimes$U(1)$_Y$, but with a new LRS transformation, that reverses both the L-R chirality and the local quantum number of a particle.  Specifically the new LRS transformation will change the $L - R$ chirality and change the gauge group representation into its charge conjugate representation, with an exception for the SU(2) representation where the $L - R$ chirality change is also changing  SU(2)$_L$ representation into SU(2)$_R$ representation and vice versa.  Since the gauge couplings are real, even though this new LRS transformation is different from the original LRS transformation, the gauge couplings of SU(2)$_L$and SU(2)$_R$ are still equal, $g_L = g_R \equiv g$.  

%\begin{center}
\begin{table}[t]
\caption{Irreducible Representation and Quantum numbers assignment for the SM and the L-R dual particles with respect to the L-R gauge group, and the global quantum number $F$.}
\centering
{\begin{tabular}{@{}ccccc@{}} \toprule
SM-particles & Irreps & & Dual-particles & Irreps \\ \hline\hline
 & & & & \\
Fermion ($F=1$) & & & & \\
$L_L \equiv \left( \begin{array}{l}
\nu \\
e
\end{array}\right)$ & (\textbf{1},\textbf{2},\textbf{1},-1) & &
$L_R \equiv \left( \begin{array}{l}
E \\
N \end{array}\right)$ &(\textbf{1},\textbf{1},\textbf{2$^*$},1) \\ 
$\nu_R$ &(\textbf{1},\textbf{1},\textbf{1},0)& & $N_L$ & (\textbf{1},\textbf{1},\textbf{1},0) \\ 
 $e_R$ & (\textbf{1},\textbf{1},\textbf{1},-2)& &$E_L$ & (\textbf{1},\textbf{1},\textbf{1},2) \\ 
$Q_L \equiv \left( \begin{array}{c}
u \\
d
\end{array}\right)$ &(\textbf{3},\textbf{2},\textbf{1},$\frac{1}{3}$) & &
$Q_R = \left( \begin{array}{c}
D \\
U
\end{array}\right)$ &(\textbf{3$^*$},\textbf{1},\textbf{2$^*$},$\frac{-1}{3}$)\\
$u_R$ & (\textbf{3},\textbf{1},\textbf{1},$\frac{4}{3}$)& & $U_L$ & (\textbf{3$^*$},\textbf{1},\textbf{1},$\frac{-4}{3}$)\\ 
$d_R$ & (\textbf{3},\textbf{1},\textbf{1},$\frac{-2}{3}$)& &
  $D_L$ & (\textbf{3$^*$},\textbf{1},\textbf{1},$\frac{2}{3}$) \\
			& & & & \\\hline 
		& & & & \\
		Scalar ($F=0$) & & & & \\
	$\chi_L = \left( \begin{array}{c}
\chi_\nu \\
\chi_e
\end{array}\right)$ &(\textbf{1},\textbf{2},\textbf{1},-1)& &
$\chi_R = \left( \begin{array}{c}
\chi_E \\
\chi_N
\end{array}\right)$ &(\textbf{1},\textbf{1},\textbf{2$^*$},1)\\
 $\eta$ & (\textbf{3},\textbf{1},\textbf{1},$\frac{4}{3}$)& &$\eta^*$ & (\textbf{3$^*$},\textbf{1},\textbf{1},$\frac{-4}{3}$) \\ 
 $\rho$ & (\textbf{3},\textbf{1},\textbf{1},$\frac{-2}{3}$)& &$\rho^*$ & (\textbf{3$^*$},\textbf{1},\textbf{1},$\frac{2}{3}$) \\ 
	& & & & \\\hline\hline
\end{tabular}\label{tabel1} }
\end{table}
%\end{center}

We add an additional global quantum number $F$ whose value is one for all fermions (minus one for anti fermions) and zero for all scalars and gauge bosons.  We assume that this global quantum number $F$ is not affected by the new LRS transformation, and therefore the L-R dual particles are different from the SM antiparticles.  The fermion and scalar content of the model are given in Table \ref{tabel1}, where we have classified the particles into two categories: the usual SM particles (plus the right-handed singlet neutrinos), and the corresponding L-R dual particles.  These two class of particles are related by the new L-R symmetry transformation.  Since both left and right chiral particles exist, this model is free from chiral anomaly. 

The particle content in this new model is a variant of the one proposed around two decades ago by Coutinho et.al. \cite{coutinho}.  However, due to the new L-R symmetry transformation and the global quantum number $F$, unlike in \cite{coutinho}  there is no need of a seesaw mechanism for generating charged fermion masses (no need of double seesaw for neutrinos) in this model.  But due to the global quantum number $F$, there is no Majorana neutrinos and therefore no leptogenesis in this model.  The L-R dual particles are very massive but they cannot decay into the SM particles via any gauge interactions. Therefore, we introduce two leptoquarks $\rho$ and $\eta$ with their corresponding L-R duals to the model as facilitators for the decay of the massive L-R dual particles into the SM particles and the neutrinos.  These two leptoquarks and their L-R duals will also produce a baryon asymmetric universe (BAU) for the SM and the L-R dual sectors. Thus the existence of the leptoquarks are necessary in this model, not only for facilitating the decay of L-R dual particle but also for producing BAU. 
%The particle content in this new model is a variant of the one in \cite{coutinho}.  However, due to the new L-R symmetry transformation and the global quantum number $F$, unlike in \cite{coutinho}  there is no need of a seesaw mechanism for generating charged fermion masses (no need of double seesaw for neutrinos) in this model.  But due to the global quantum number $F$, there is no Majorana neutrinos and therefore no leptogenesis in this model.  Consequently, we have to introduce two leptoquarks $\rho$ and $\eta$ with their corresponding L-R duals, not only to produce a baryon asymmetric universe but also to act as facilitators for the decay of the massive L-R dual particles into the SM particles and the neutrinos.

\section{Scalar Sector}
The scalar potential that is invariant under the gauge group and the new L-R transformation is given by the following,
\begin{align}
\mathcal{V} =  -&\mu_L^2 |\chi_L|^2 -\mu_R^2|\chi_R|^2 
+\mu_1^2 |\eta|^2 +\mu_2^2 |\rho|^2  + \lambda_1\left(|\chi_L|^4 + |\chi_R|^4   \right) +  \lambda_2 |\eta|^4 +\lambda_3 |\rho|^4  \nonumber\\ +& \epsilon_1 |\eta|^2\left(|\chi_L|^2+|\chi_R|^2\right)  
 +  \epsilon_2 |\rho|^2\left(|\chi_L|^2+|\chi_R|^2\right) + \epsilon_3 |\eta|^2|\rho|^2 + \epsilon_4 |\chi_L|^2|\chi_R|^2 \nonumber\\
+& \epsilon_5 \rho^\dagger \eta \eta^\dagger \rho .
\end{align} 
The parameter $\mu_{L}\neq \mu_{R}$ corresponds to a soft L-R symmetry breaking term. The parameters
$\mu_i$'s, $\lambda_i$'s and $\epsilon_i$'s above can be chosen so that the scalar potential can lead to non zero vacuum expectation values (VEVs) for certain scalars.   The leptoquarks $\eta$ and $\rho$ should not have any non zero VEV, since otherwise the gluons will be massive.  The $\chi_L$ and $\chi_R$ can have different VEV's, written as follows,
\begin{equation}
\langle \chi_L\rangle = \left( \begin{array}{c} v_L \\
0
\end{array}\right);\quad \langle \chi_R\rangle = \left( \begin{array}{c}
0 \\
v_R
\end{array}\right) ,
\end{equation} 
where in general $v_L \neq v_R$.  These non zero VEV's for $\chi_L$ and $\chi_R$ will give masses to the left and right weak gauge bosons.  Note that because there is no bidoublet scalar in this model then there is no direct mixing between $W_L$ and $W_R$.

The relevant terms for the mass of the weak gauge bosons, are as follows,
\begin{align}
&\left|\sqrt{\frac{1}{2}}\left(- \frac{i}{2} g {\tau} \cdot \textbf{W}_L + \frac{i}{2} g' B_\mu\right) \left( \begin {array}{c} v_L\\\noalign{\medskip}0\end {array} \right) \right|^2+\left|\sqrt{\frac{1}{2}} \left(-\frac{i}{2}g\tau \cdot \textbf{W}_R - g'\frac{i}{2}B_\mu\right)\left( \begin {array}{c} 0\\\noalign{\medskip}v_R\end {array} \right) \right|^2\nonumber \\ 
=&\left(\frac{1}{2}gv_L\right)^2W^+_{\mu L}W^{\mu -}_L+\left(\frac{1}{2}gv_R\right)^2W^+_{\mu R}W^{\mu -}_R+\frac{1}{2}\left(W^3_{\mu L} \ \ W^3_{\mu R} \ \ B_\mu\right)\textbf{M}_{\rm WB}\left(\begin {array}{c} W^{3\mu}_L\\\noalign{\medskip}W^{3\mu}_R\\\noalign{\medskip}B_\mu\end {array}\right),
\label{wlr}
\end{align}
where $\textbf{M}_{\rm WB}$ is the mass matrix of $W^{3\mu}_{L}, W^{3\mu}_{R}$ and $B^\mu$ fields which is given by,
\begin{equation}
\textbf{M}_{\rm WB}=\frac{v_R^2}{4}\bordermatrix{& & &\cr
	&g^2\omega^2 &0 &-gg'\omega^2\cr
	&0 &g^2 &-gg'\cr
	&-gg'\omega^2 &-gg' &g'^2(1+\omega^2)\cr} ,
\label{wlwrB}
\end{equation}
where $\omega = v_L/v_R$. The first and second terms in Eq.\eqref{wlr} directly give the mass of charged gauge bosons, $W_L$ and $W_R$, namely, 
\begin{equation}
M_{W_L}^2 =\frac{1}{4}g^2 v_L^2, \quad  M_{W_R}^2 = \frac{1}{4}g^2 v_R^2.  
\end{equation}
While diagonalizing the matrix $\textbf{M}_{\rm WB}$ in Eq.\eqref{wlr}, we obtain the masses of  the neutral right and left weak gauge bosons ($Z_R,Z_L$) and the photon ($A$), 
\begin{align}
M_{Z_R}^2  = & \frac{v_R^2}{8}\left(g^2+g'^2\right)\left((1+\omega^2) + \left((1-\omega^2)^2 + \frac{4 \omega^2 g'^4}{(g^2+g'^2)^2} \right)^{1/2} \right)\nonumber\\
\simeq & v_R^2 g^2 \frac{\tan^2 \theta_W}{\sin^2 2\beta} \left( 1 + \omega^2 \sin^4 \beta\right), \\
M_{Z_L}^2  = & \frac{v_R^2}{8}\left(g^2+g'^2\right)\left((1+\omega^2) - \left((1-\omega^2)^2 + \frac{4 \omega^2 g'^4}{(g^2+g'^2)^2} \right)^{1/2} \right)\nonumber\\
\simeq & \frac{1}{4}\frac{v_L^2 g^2}{\cos^2 \theta_W}\left(1 - \omega^2 \sin^4 \beta \right), \\
M_A^2 =& 0, 
\end{align}
where the mixing angles $\theta_W$ and $\beta$ are given by,
\begin{equation}
\sin^2 \theta_W = \frac{g^2 {g'}^2}{g^4 + 2 g^2{g'}^2}, \quad \sin^2 \beta = \frac{{g'}^2}{g^2 + {g'}^2}.
\end{equation}
The phenomenology of our model is similar to many LRS models. In particular, it is the same as in \cite{coutinho} where we have reproduced some of their result here. 
The mass basis weak gauge bosons are related to the original gauge bosons as,
\begin{align}
A_\mu  = & \sin \theta_W \ W_{\mu L}^3 + \sin \theta_W \ W_{\mu R}^3 + \cos \beta \cos \theta_W \ B_\mu,\label{relA} \\
Z_{\mu L} \simeq & -\cos \theta_W \ W_{\mu L}^3 + \sin \theta_W \sin \beta \  W_{\mu R}^3 + \sin \theta_W \cos \beta \ B_\mu,\\
Z_{\mu R} \simeq & -\omega^2\sin^2 \beta \cos \beta \ W_{\mu L}^3  - \cos \beta \ W_{\mu L}^3 + \sin \beta \ B_\mu.
\end{align} 
From Eq.\eqref{relA}, we have the usual relation for the electromagnetic charge operator $Q = T^3_L + T^3_R + Y/2$. 
The neutral currents coupled to the massive vector bosons $Z_L$ and $Z_R$ are given by,
\begin{align}
J^{\mu}_{ Z_{L}} =& \frac{g}{\cos \theta_{W}} \gamma^{\mu} \left\{ (1-\omega^{2} \sin^{4}\beta)
T^{3}_{L} - \omega^{2} \sin^{2}\beta \cos^{2}\beta \ T^{3}_{R} 
- Q \sin^{2}\theta_{W} \left( 1- \frac{\omega^{2} \sin^{4}\beta }{\sin^{2}\theta_{W} } \right)  \right\} \\
J^{\mu}_{ Z_{R}} =& g \tan \theta_{W} \tan \beta \ \gamma^{\mu} 
\left\{ \left( 1+\frac{\omega^{2} \sin^{2}\beta \cos^{2}\beta }{\sin^{2}\theta_{W} }  \right) T^{3}_{L}
+\frac{T^{3}_{R}}{\sin^{2}\beta} 
- Q(1+\omega^{2} \cos^{2}\beta \sin^{2}\beta)
\right\} \label{JzR}
\end{align}
From Eq.\eqref{JzR}, it is clear that the SM fermion's coupling to the $Z_R$ are not suppressed by a factor $\omega$. This has been used in \cite{coutinho} to calculate the corrections of the SM fermion's coupling $Z_{L}$. By fitting this correction to the experimental data, Countinho et. all. have obtained a lower bound for $v_{R}$, i.e. $v_{R}>30 v_{L} $ \cite{coutinho}.

\section{Fermion Sector}
\label{fer}
The Yukawa terms in the Lagrangian that are invariant under the gauge, and the new LRS transformation are,
\begin{eqnarray}\label{38a} 
& &- G_e(\bar l_{L}\chi_L e_{R}  + \bar L_{R}\tilde\chi_R E_{L})  - G_{\nu}(\bar l_{L}\tilde\chi_L \nu_{R} + \bar L_{R}\chi_R N_{L}) -G_{d}(\bar q_{L}\chi_L d_{R}  + \bar{Q}_{R}\tilde\chi_R D_{L}) \nonumber \\ & &- G_{u}(\bar q_{L}\tilde\chi_L u_{R} + \bar{Q}_{R}\chi_R U_{L})- G_{d\nu}(\bar{\nu}_{R} \rho  D_{L}+ \bar{N}_L\rho^\dagger d_R ) - G_{ue}(\bar U_{L}\rho^\dagger e_{R} + \bar u_{R}\rho E_{L}) \nonumber \\ & &-G_{du} (\bar{U}_{L}\rho d_{R}+\bar{u}_R\rho^\dagger D_{L}) - G_{u\nu} (\bar{U}_{L}\eta^{\dagger} \nu_{R}+\bar{u}_R \eta N_{L}) - G_{dd} \bar D_{L}\eta d_{R} - M \bar \nu_{R} N_{L} + {\rm h.c.} 
\end{eqnarray}
Note that the Yukawa couplings $G$'s and the mass $M$ are $3 \times 3$ matrices to account for the three generations.
After the $\chi_R$ and $\chi_L$ gain non zero VEV, the above Yukawa terms will generate mass for the charged fermions and mixing mass terms for the neutrinos,
\begin{eqnarray}
& & -G_{d}(v_L\bar d_{L} d_{R}  + v_R\bar D_{R} D_{L}) - G_{u}(v_L\bar u_{L} u_{R} + v_R \bar U_{R}U_{L}) -G_e(v_L\bar e_{L} e_{R}  + v_R \bar E_{R} E_{L}) \nonumber \\ & & - G_{\nu}(v_L\bar \nu_{L} \nu_{R} + v_R\bar N_{R}N_{L}) -  M \bar \nu_{R} N_{L} + {\rm h.c.}
\label{lagql}
\end{eqnarray}
The mass of the L-R dual fermions (except neutrinos) will be $\omega^{-1}$ times the mass of its corresponding SM particles. The current limit on the charged lepton mass is around 100 GeV \cite{pdg}.  This means we should have $v_R > 10^5  v_L$.  This bound surpasses the bound obtained by Coutinho et. al. \cite{coutinho}  from the correction of the $Z_{L}$ coupling to the SM fermion, that has been mentioned in the previous section. 
For the case of neutrinos,  in general the value of $M$ in Eq.\eqref{lagql} is not restricted by any gauge symmetry mechanism since $\nu_{R}$, $N_{L}$ are gauge singlet, thus we assume that $M \gg v_L,v_R$.  This will lead to a seesaw mechanism~\cite{seesaw}.  The terms in Eq.\eqref{lagql} related to the neutrino masses can be written as $\bar{\psi}\mathcal{M}\psi$ where $\psi=\left(\nu_{L},N_{R}, N_{L},\nu_{R}\right)^T $ and 
\begin{equation}
\mathcal{M} = \begin{pmatrix} 
 0  & 0  & 0  & G_\nu v_L   \\
 0 & 0  & G_\nu v_R  & 0  \\
 0  & G_\nu v_R  & 0  & M  \\
 G_\nu v_L   & 0  & M & 0 
\end{pmatrix}, \label{matrix}
\end{equation}
which is similar in form to a Dirac seesaw mechanism case in \cite{ronca}. From this it is clear that in our model the neutrinos are Dirac particles. Therefore, any positive result from the neutrinoless double beta decay will rule out this model. 
The mass matrix in Eq.\eqref{matrix} can be diagonalized and written in terms of mass basis neutrino state $\Psi = \left(\nu, \nu', N', N\right)$, where $\psi = \mathcal{U} \Psi$, with the mixing matrix $\mathcal{U}$ diagonalizes $\mathcal{M}$, i.e. $\mathcal{M} = \mathcal{U}\mathcal{M}_\delta\mathcal{U}^T$.  Using the seesaw mechanism, we have
\begin{equation}
\mathcal{U} \simeq \begin{pmatrix}
I/\sqrt{2} & I/\sqrt{2}  &  v_L G_\nu M^{-1} & -v_L G_\nu M^{-1}\\
-I/\sqrt{2}  & I/\sqrt{2}  & v_R G_\nu M^{-1} & v_R G_\nu M^{-1}  \\
-M^{-1}v_L G_\nu^T & -M^{-1}v_R G_\nu^T  & I/\sqrt{2}  & -I/\sqrt{2}  \\
M^{-1}v_L G_\nu^T & - M^{-1}v_R G_\nu^T  & I/\sqrt{2}  & I/\sqrt{2}  
\end{pmatrix}\mathcal{V},\label{Umatriks}
\end{equation}
with $I$ is a $3\times 3$ identity matrix, $\mathcal{V}$ is the following block diagonal matrix,
\begin{equation}
\mathcal{V} \simeq \begin{pmatrix}
V_1 & 0  &  0 & 0 \\
0  & V_2 & 0 & 0  \\
0 & 0  & V_3  & 0 \\
0 & 0 & 0  & V_4  
\end{pmatrix},\label{Vmatriks}
\end{equation}
where  the $3\times 3$ matrices $V_i$ ($i = 1,\dots,4)$ are the matrix that will diagonalize each block sub matrix in the following diagonal matrix $\mathcal{M}_\delta$,
\begin{equation}\label{diagmass}
\mathcal{M}_\delta \simeq \mathcal{V}^T\begin{pmatrix}
v_R v_L  G_\nu M^{-1} G_\nu^T & 0 & 0 & 0\\
 0  & - v_R v_L  G_\nu M^{-1}  G_\nu^T  & 0 & 0  \\
 0 & 0  & M  & 0  \\
0 & 0  & 0  & -M 
\end{pmatrix}\mathcal{V}.
\end{equation}
For example $V_1$ will diagonalize $v_R v_L  G_\nu M^{-1} G_\nu^T$.  Because the first two block matrices are the same, then $V_1 = V_2$.  Similarly for the next two block matrices, thus we have $V_3 = V_4$.  From Eq.\eqref{diagmass}, we have two different orders for the mass eigenvalues, 
\begin{equation}
m_{\nu} = m_{\nu'} \simeq  v_L v_R V_1^T G_{\nu} M^{-1} G_{\nu}^T V_1 , \quad m_{N} = m_{N'}   \simeq V_3^T M V_3.
\label{massanu}
\end{equation}
The doublets neutrinos ($\nu_L$ and $N_R$) are dominated by lighter neutrinos $\nu$ and $\nu'$, while the singlet neutrinos ($\nu_R$ and $N_L$) are dominated by heavier neutrinos $N$ and $N'$.  But the $\nu_L, N_R$ still contain a small portion of  heavier component neutrinos, while the $\nu_R, N_L$ still contain a small portion of  lighter component neutrinos.   We can have an estimation for the mass order of $M$.  Assuming that the light neutrinos have masses in the range of  $\sqrt{\Delta m_{12}^2}$ to $\sqrt{\Delta m_{23}^2}$ as in the neutrino oscillation results \cite{pdg}, then the order of the light neutrino is $m_{\nu} \simeq 10^{-12} \sim 10^{-11}$ GeV.  If we assume that the value of $G_\nu$ is of the same order as the Yukawa coupling for charged lepton, i.e. $G_e\simeq 10^{-6} \sim 10^{-3}$ then $M$ should be in the order of $ (10^{5} \sim 10^{12}) \omega^{-1}$ GeV.

There is no oscillation between SM doublet neutrinos and L-R dual doublet neutrinos. This can be seen clearly if we write down the flavor basis neutrinos in terms of the mass basis neutrinos.  For SM doublet neutrinos and its L-R duals, we have
\begin{eqnarray}
\nu_{L\alpha} &\simeq & (V_1/\sqrt{2})_{\alpha i} \nu_i +  (V_2/\sqrt{2})_{\alpha i} \nu'_i +v_L (G_\nu M^{-1} V_3)_{\alpha i} N'_i - v_L (G_\nu M^{-1} V_4)_{\alpha i} N_i \nonumber\\
N_{R\beta} &\simeq & - (V_1/\sqrt{2})_{\beta j} \nu_j +  (V_2/\sqrt{2})_{\beta j} \nu'_j +v_R (G_\nu M^{-1} V_3)_{\beta j} N'_j + v_R (G_\nu M^{-1} V_4)_{\beta j} N_j,
\end{eqnarray}
where $\alpha,\beta$ are the flavour index and the sum in $i$ and $j$ are over $1,2,3$.  From this, the probability amplitude for a SM doublet neutrino $\nu_{L\alpha}$ with energy $E$ oscillates into an L-R dual doublet neutrino $N_{R\beta}$ after traveling a distance $L$ is 
\begin{eqnarray}
\langle N_{R\beta}|\nu_{L\alpha} \rangle &\simeq& - \frac{1}{2}(V_1)_{\beta i} (V_1)^{*}_{\alpha i} \exp\left(-i\frac{m_{\nu_i}^2}{2E} L\right)  + \frac{1}{2} (V_2)_{\beta i} (V_2)^{*}_{\alpha i} \exp\left(-i\frac{m_{\nu_i'}^2}{2E} L\right) \nonumber\\ &+&  v_L v_R (G_\nu M^{-1} V_3)_{\beta i}(G_\nu M^{-1} V_3)^{*}_{\alpha i} \exp\left(-i\frac{m_{N'_i}^2}{2E} L\right) \nonumber\\ &-&   v_L v_R (G_\nu M^{-1} V_4)_{\beta i}(G_\nu M^{-1} V_4)^*_{\alpha i} \exp\left(-i\frac{m_{N_i}^2}{2E} L\right) .
\end{eqnarray}
It is clear that the above amplitude is approximately zero due to $V_1=V_2$, $V_3=V_4$, $m_\nu = m_{\nu'}$, and $m_N=m_{N'}$.  The probability amplitude for a SM doublet neutrino $\nu_{L\alpha}$ with energy $E$ oscillates into a SM doublet neutrino $\nu_{L\beta}$ after traveling a distance $L$ is
\begin{eqnarray}
\langle \nu_{L\beta}|\nu_{L\alpha} \rangle &\simeq& \frac{1}{2} (V_1)_{\beta i} (V_1)^{*}_{\alpha i} \exp\left(-i\frac{m_{\nu_i}^2}{2E} L\right)  +  \frac{1}{2}(V_2)_{\beta i} (V_2)^{*}_{\alpha i} \exp\left(-i\frac{m_{\nu'_i}^2}{2E} L\right) \nonumber\\ &+&  v_L^2 (G_\nu M^{-1} V_3)_{\beta i}(G_\nu M^{-1} V_3)^{*}_{\alpha i} \exp\left(-i\frac{m_{N'_i}^2}{2E} L\right) \nonumber \\ &+&   v_L^2 (G_\nu M^{-1} V_4)_{\beta i}(G_\nu M^{-1} V_4)^*_{\alpha i} \exp\left(-i\frac{m_{N_i}^2}{2E} L\right) \nonumber\\
&\simeq& (V_1)_{\beta i} (V_1)^{*}_{\alpha i} \exp\left(-i\frac{m_{\nu_i}^2}{2E} L\right),
\end{eqnarray}
where we have neglected the small terms of order $|v_L G_\nu M^{-1}|^2$.  This last result will lead to the usual formulation for the three SM neutrinos oscillation.  The oscillation probability between doublet neutrinos (either the SM or its L-R dual) and singlet neutrinos is very small, the largest being of order $|v_R G_\nu M^{-1}|^2$.

\begin{figure}[t]
	\centering
		\includegraphics[width=.5\linewidth]{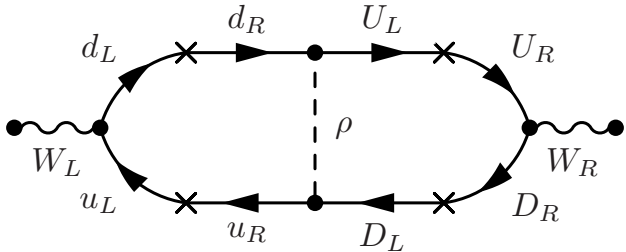}
	\caption{The two-loop $W_{L}-W_{R}$ mixing. }
\label{twoloop}
\end{figure}

There is no mixing between the SM and L-R dual fermions in this model, thus we do not have a $W_{L} - W_{R}$ mixing at a one-loop level which usually appears in the conventional LRS models. Instead, the SM and L-R dual particles (leptons and quarks) via the leptoquark $\rho$ can mediate the $W_{L} - W_{R}$ mixing at two-loop level. The highest contribution for the two-loop $W_{L} - W_{R}$ mixing is shown in Figure \ref{twoloop}, where the approximate two-loop mass term is given by, 
\begin{align}
\delta M^{2}_{W_{L} W_{R}} \simeq& \frac{3 g^{2} |V_{tb}|^2 G_{ud}^2}{2(16\pi^{2})^{2}} \frac{M_{T} M_{B} m_{t} m_{b}}{M_{\rho}^{2} } \nonumber \\
\simeq & \frac{3 g^{2} |V_{tb}|^2 G_{ud}^2}{2(16\pi^{2})^{2}} m_{b}^{2},
\end{align} 
where $V$ is the Cabibbo-Kobayashi-Maskawa matrix \cite{cabibo}. $M_{T}$ and $M_{B}$ are the mass of top and bottom L-R dual quarks, respectively. From the above equation, it is clear that if $M_{T}\ll M_{\rho}$ then the $W_{L} - W_{R}$ mixing is a negligible quantity. 

\section{The Leptoquarks}
%The Lagrangian of the model is invariant under global quantum number $F$ so there are no massive Majorana neutrinos whose decay can lead to leptogenesis.  But  baryon asymmetry can still occur because of baryogenesis from the decay of leptoquarks.  The leptoquarks are also needed in this model for facilitating the decay of massive L-R dual particles into  SM particles.  In this section, we will first discuss the baryogenesis and then later how the leptoquarks can facilitate the decay of the L-R dual particles.  

%The existence of leptoquarks in a model may lead to baryon number violation \cite{weinberg3} which may lead to baryogenesis if the  Sakharov's conditions \cite{sakharov} are satisfied, namely (i) baryon number violation, (ii) C and CP violations and (iii) a departure from thermal equilibrium.  In our model, the decay of the two leptoquarks $\rho$ and $\eta$ is the same as the textbook case in \cite{kolbturner}.   The leptoquarks can decay into both the SM fermions and their L-R duals. The decays of leptoquarks, up to one-loop order that can lead to baryogenesis, are shown in Figure \ref{figure1}-\ref{figure4}.  We summarize the value of the baryon number production for those decays in Table \ref{table1}.

\begin{figure}[t]
	\centering
	\begin{minipage}{.5\textwidth}
		\centering
		\includegraphics[width=.9\linewidth]{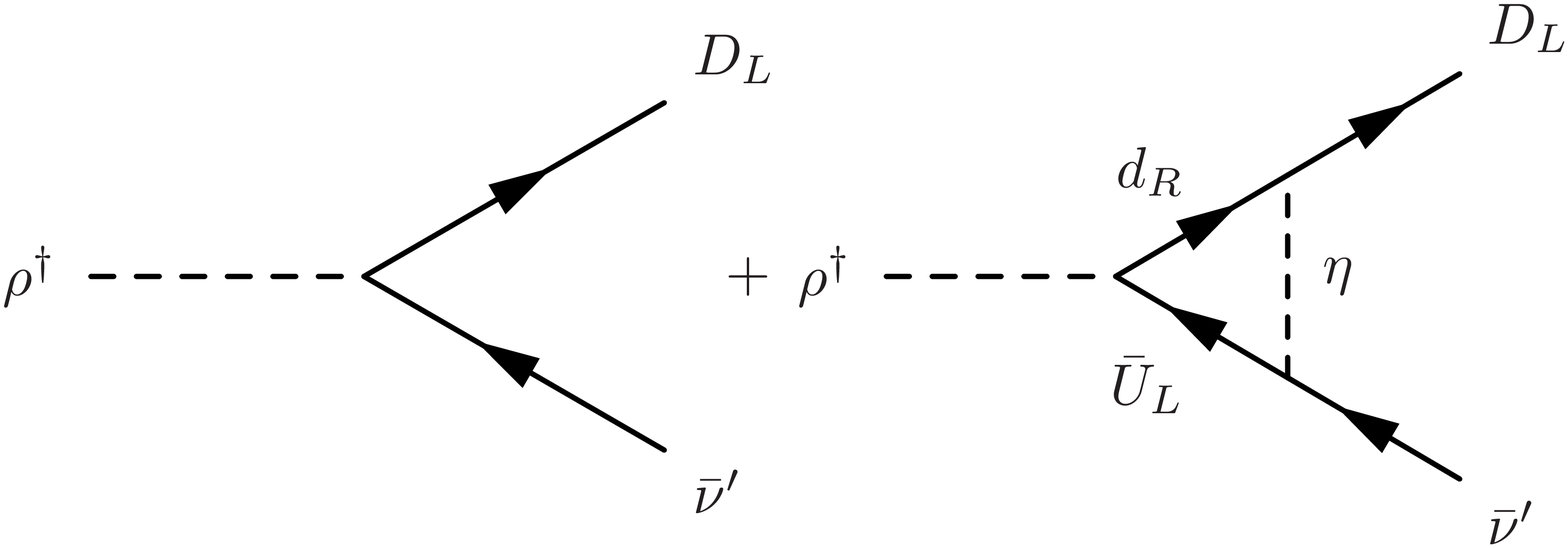}
		\caption{Process  $\rho^{\dagger}\rightarrow D_{L} + \bar{\nu}^{\prime}$}
		\label{figure1}
	\end{minipage}%
	\begin{minipage}{.5\textwidth}
		\centering
		\includegraphics[width=.9\linewidth]{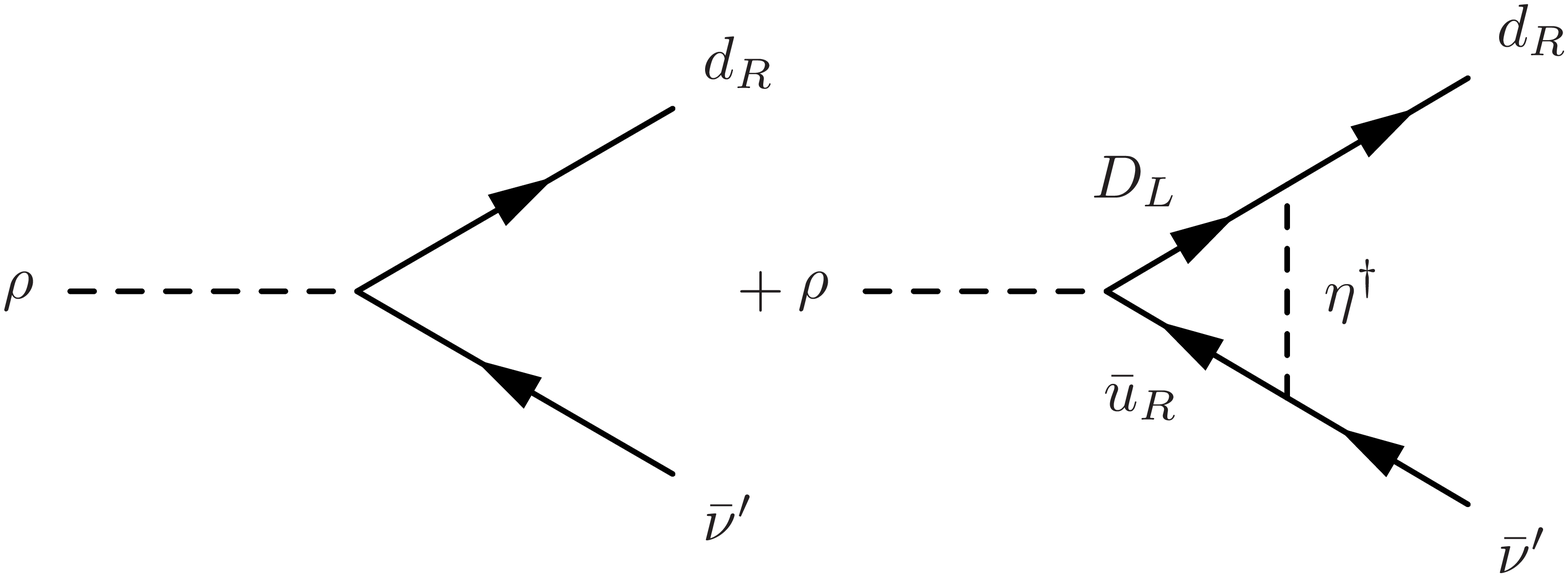}
		\caption{Process  $\rho\rightarrow d_{R} + \bar{\nu}^{\prime}$}
		\label{figure2}
	\end{minipage}
	\\
%\end{figure}
%\begin{figure}[t]
	\centering
	\begin{minipage}{.5\textwidth}
		\centering
		\includegraphics[width=.9\linewidth]{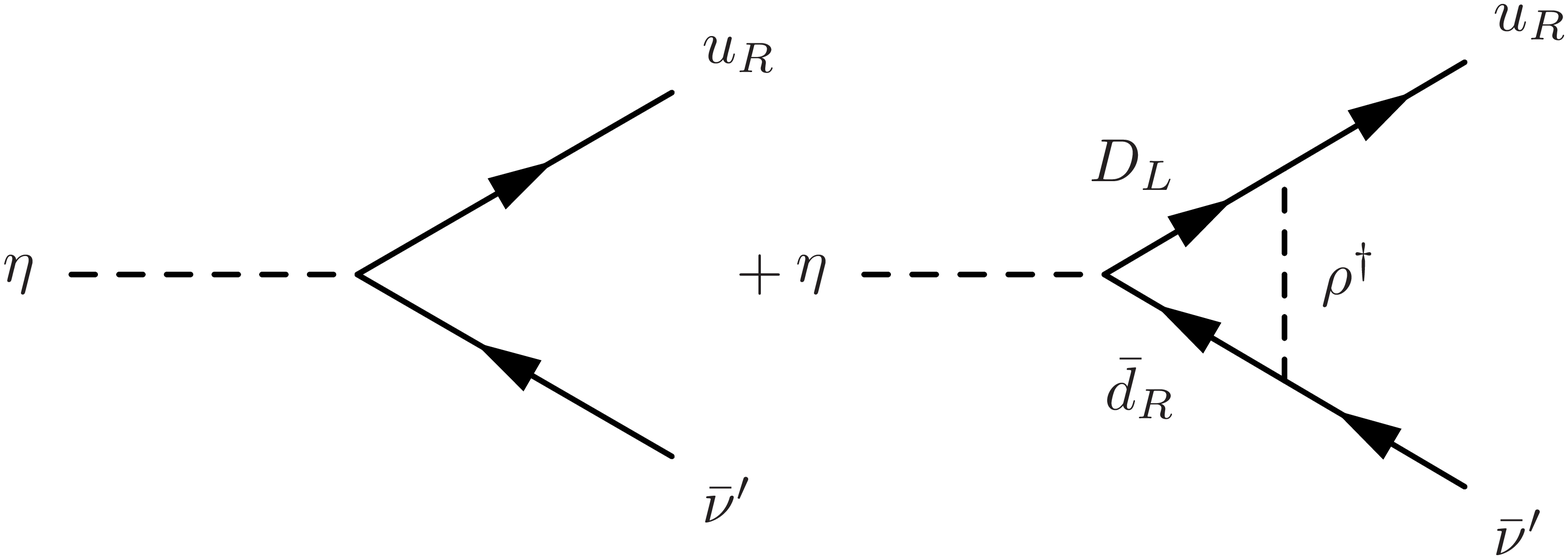}
		\caption{Process  $\eta\rightarrow u_{R} + \bar{\nu}^{\prime}$}
		\label{figure3} 
	\end{minipage}%
	\begin{minipage}{.5\textwidth}
		\centering
		\includegraphics[width=.9\linewidth]{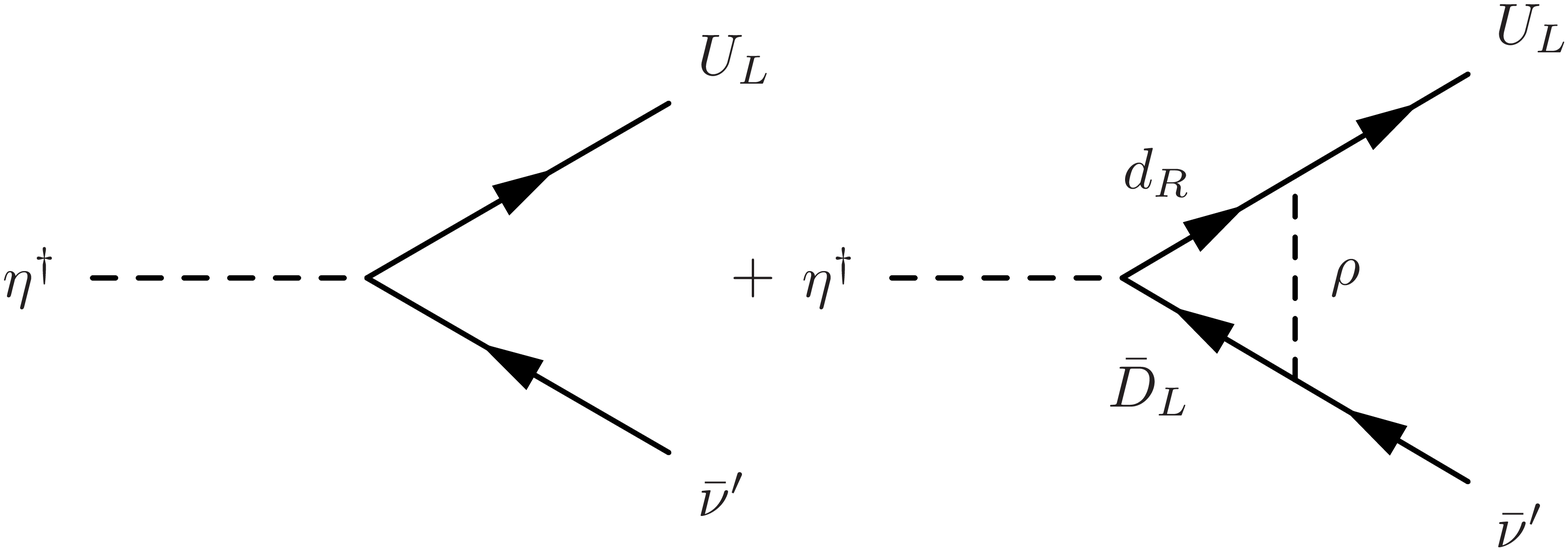}
		\caption{Process  $\eta^{\dagger}\rightarrow U_{L} + \bar{\nu}^{\prime}$}
		\label{figure4}
	\end{minipage}
\end{figure}

The Lagrangian of the model is invariant under global quantum number $F$ so there are no massive Majorana neutrinos whose decay can lead to leptogenesis.  But  baryon asymmetry can still occur because of baryogenesis from the decay of leptoquarks. The existence of leptoquarks in a model may lead to baryon number violation \cite{weinberg3} which may lead to baryogenesis if the  Sakharov's conditions \cite{sakharov} are satisfied.  The calculation of the baryon number production from the decay of the leptoquarks follows the usual method as in \cite{kolbturner}, only here we have to consider for the leptoquarks and its L-R duals with the decay results are the SM and the L-R dual particles.  The decays of leptoquarks up to one-loop order that can lead to baryogenesis are shown in Figure \ref{figure1}-\ref{figure4}.  We summarize the value of the baryon number production for those decays in Table \ref{table1}. 

\begin{table}
	\centering
	\caption{The branching for decay processes of the leptoquark particles and their baryon number violations ($ \Delta B$).}
	\label{table1}       % Give a unique label
	\begin{tabular}{lllll}
		\hline\hline\noalign{\smallskip}
		Particles &  & Final state &$ \Delta B$ \\
		\noalign{\smallskip}\hline\noalign{\smallskip}
		$\rho^{\dagger}$ & $\longrightarrow$ & $D_{L} + \bar{\nu'}$  & $-1/3$  \\ 
		
		$\rho$ & $\longrightarrow$ & $\bar{D}_{L} + \nu'$  & $+1/3$  \\ 
		
		$\rho$ & $\longrightarrow$ & $d_{R} + \bar{\nu'}$  & $+1/3$  \\ 
		
		$\rho^{\dagger}$ & $\longrightarrow$ & $\bar{d}_{R} + \nu'$  & $-1/3$  \\ 
		
		$\eta$ & $\longrightarrow$ & $u_{R} + \bar{\nu'}$ & $+1/3$  \\ 
		
		$\eta^{\dagger}$ & $\longrightarrow$ & $\bar{u}_{R} + \nu'$   & $-1/3$  \\ 
		
		$\eta^{\dagger}$ & $\longrightarrow$ & $U_{L} + \bar{\nu'}$ & $-1/3$  \\ 
		
		$\eta$ & $\longrightarrow$ & $\bar{U}_{L} + \nu'$ & $+1/3$  \\ 
		\noalign{\smallskip}\hline\hline
	\end{tabular}
\end{table}

The mean net baryon number produced in the decay of a particle $X$ is given by \cite{kolbturner},
\begin{align}
\epsilon_{X} = \sum_{f}B_{f}\frac{\Gamma(X\rightarrow f)-\Gamma(\bar{X}\rightarrow \bar{f})}{\Gamma_{X}} , \label{epsCP}
\end{align}
where the sum $\sum_f$ runs over all final state fermion $f$, $B_{f}$ is the produced baryon number of the decay process and $\Gamma_{X}$ is the total decay rate of $X$.  The contribution to $\epsilon$ comes from the interference between the lowest (tree-level) order and the one-loop order diagrams, as shown in Fig. \ref{figure1}-\ref{figure4}.  From Fig. \ref{figure1}, the decay rate difference between the decay $\rho^{\dagger}\rightarrow D_{L} + \bar{\nu}^{\prime}$ and $\rho\rightarrow \bar{D}_{L} + \nu^{\prime}$ is
\begin{align}
\Gamma(\rho^{\dagger}\rightarrow D_{L} + \bar{\nu'})-\Gamma(\rho\rightarrow \bar{D}_{L} + \nu') = -4\ |v_R G_\nu M^{-1}|^2{\rm Im}\left(I^{(1)}_{\rho\eta}\right)\ {\rm Im} (G_{d\nu}G_{du}^{*}G_{dd}G_{u\nu}^{*}), \label{decayrate01}
\end{align} 
where $(I^{(1)}_{\rho\eta})$ is the kinematic factor of the internal loop in Fig. \ref{figure1} due to the exchange of $\eta$ in the decay of $\rho^{\dagger}$. As for the decay rate difference between $\rho\rightarrow d_{R} + \bar{\nu}^{\prime}$ and $\rho^{\dagger}\rightarrow \bar{d}_{R} + \nu^{\prime}$ (Fig. \ref{figure2}), we have,
\begin{align}
\Gamma(\rho\rightarrow d_{R} + \bar{\nu'})- \Gamma(\rho^{\dagger}\rightarrow \bar{d}_{R} + \nu') = -4\ |v_R G_\nu M^{-1}|^2{\rm Im}\left(I^{(2)}_{\rho\eta}\right)\ {\rm Im} (G_{dd}^{*}G_{du}G_{u\nu}G_{d\nu}^{*}), \label{decayrate02} 
\end{align}
where, similar as before, $(I^{(2)}_{\rho\eta})$ is the kinematic factor of the internal loop in Fig. \ref{figure2} due to the exchange of $\eta^{\dagger}$ in the decay of $\rho$. Both ${\rm Im}(I^{(1)}_{\rho\eta})$ and ${\rm Im}(I^{(2)}_{\rho\eta})$ in Eqs.\eqref{decayrate01} and \eqref{decayrate02} have the same value and they are given by,
\begin{eqnarray}
{\rm Im}\left(I^{(1)}_{\rho\eta}\right) = {\rm Im}\left(I^{(2)}_{\rho\eta}\right) = \frac{M_{\rho}}{128\pi^{2}}\left\{1-\sigma^{2}\ln\left(1+\sigma^{-2}\right)\right\} ,
 \label{decayrate03} 
\end{eqnarray}
where $\sigma :=M_{\rho}/M_{\eta}$, while $M_{\rho}$ and $M_{\eta}$ are the mass of the leptoquarks $\rho$ and $\eta$, respectively. 
Substituting Eqs.\eqref{decayrate01} and \eqref{decayrate02} into Eq.\eqref{epsCP}, we obtain,
\begin{eqnarray}
\epsilon_{\rho} = \frac{1}{3\pi} \frac{{\rm Im}(G_{d\nu}G_{du}^{*}G_{dd}G_{u\nu}^{*})}{|G_{d\nu}|^{2}} \left\{1-\sigma^{2}\ln\left(1+\sigma^{-2}\right)\right\}, \label{epslain01}
\end{eqnarray}
where we have used,   
\begin{equation}\label{decaytot}
\Gamma_{\rho} \simeq |v_R G_\nu M^{-1}|^2 \frac{M_{\rho}|G_{d\nu}|^{2}}{8\pi}.
\end{equation}
Following the same steps above, we obtain the mean net baryon number produced from the decay of $\eta$ as,
\begin{eqnarray}
\epsilon_{\eta} = \frac{1}{3\pi} \frac{{\rm Im}\left(G_{dd}^{*}G_{du}G_{d\nu}G_{u\nu}^{*}\right)}{|G_{u\nu}|^{2}} \left\{1-\sigma^{-2}\ln\left(1+\sigma^{2}\right)\right\}. \label{epslain02}
\end{eqnarray}
Therefore the total mean net baryon asymmetry is given by,
\begin{align}
\epsilon = \frac{{\rm Im}(G_{d\nu}G_{du}^{*}G_{dd}G_{u\nu}^{*})}{3\pi}\left[\frac{\left\{1-\sigma^{2}\ln\left(1+1/\sigma^{2}\right)\right\}}{|G_{d\nu}|^{2}}-\frac{\left\{1-\sigma^{-2}\ln\left(1+\sigma^{2}\right)\right\}}{|G_{u\nu}|^{2}}\right].  \label{epstotal}
\end{align}
%For any value of $\sigma$, the value in each of the curly bracket above is of order unity. Thus the order of the total mean net baryon asymmetry is mainly determined by the coupling constant $G$'s which have to be complex to accommodate the C and CP violations.
For any value of $\sigma$, the value in each of the curly bracket above is of order unity. Thus the order of the total mean net baryon asymmetry is mainly determined by the coupling constant $G$'s which have to be complex to accommodate the C and CP violations.  It is important to note that the value of $\epsilon$ will vanish if the new LRS symmetry transformation does not reverse the local quantum number of particles (check the value of $\Delta B$ in Table \ref{table1}). 

The leptoquark starts to become non-relativistic at a temperature comparable to their mass $T \simeq M_{lq}$, where $M_{lq}$ is the mass of either $\rho$ or $\eta$ which we assume to be of the same order.  At this temperature, the leptoquark number density is determined primarily by their decay rate $\Gamma_D$ and the cosmic expansion rate $H \simeq g_*^{1/2}T^2/m_{Pl}$, where $g_*$ is the effective relativistic degree of freedoms and $m_{Pl}$ is the Planck mass.  If at $T \simeq M_{lq}$ the leptoquark decay rate is very small compared to the cosmic expansion rate $\Gamma_D \ll H$ then the leptoquark can become overabundant.  From this requirement and the result in Eq.\eqref{decaytot} for the leptoquark total decay rate, we have the lower bound value for the mass of the leptoquark,
\begin{eqnarray}
M_{lq} \gg \alpha_{lq}g_{*}^{-1/2}m_{Pl} |v_R G_\nu M^{-1}|^2,    \label{massLQ}
\end{eqnarray}
where $\alpha_{lq}=|G_{i}|^{2}/4\pi$ ($i=d\nu,u\nu,dd,du$).   When the leptoquarks start to decay, most of the particles (including the L-R duals) are still relativistic.  The leptoquarks contribute $g_{lq} = 2\times 2\times 3$ relativistic degree of freedoms (due to $\rho$ and $\eta$, and their antiparticles, where each has colors) to the $g_*$.  We will take the value of $g_{*}$ around 200 when the leptoquarks decay. The baryon asymmetry as the ratio of the total baryon number to the total entropy density produced by the leptoquarks decay is given by, 
\begin{eqnarray}
B &=& \frac{n_{lq}(T_{M})\epsilon}{s(T_{M})} = \frac{45\zeta(3)}{2\pi^{4}}\left(\frac{g_{lq}}{g_{*}}\right)\epsilon \nonumber \\
&\simeq & 0.016\ \epsilon, \label{BASfinal}
\end{eqnarray}
where $\zeta(n)$ is the Riemann-Zeta function and $T_M \simeq M_{lq}$.  This baryon number will remain as long as there is no more baryon number producing reaction and there is no more entropy production.  To get the current observed value of $B \simeq 10^{-10}$ \cite{baryonnumber}, the value of $\epsilon$ should be around $10^{-8}$.  From Eq.\eqref{epstotal}, the order of $\epsilon$ depends on the imaginary component of $G_{d\nu}G_{du}^{*}G_{dd}G_{u\nu}^{*}$ and the absolute value of $|G_{d\nu}|^2$ or $|G_{u\nu}|^2$.  These coupling constants may have a very small imaginary part and a larger real part.  Taking the real part of the $G_i$'s to be of order unity, then the imaginary part should be of order $10^{-8}$ to get the correct value for the $\epsilon$.    Using Eq.\eqref{massLQ} and taking $\alpha_{lq} \sim |G_{lq}|^2 \sim 1$, $|G_\nu v_R| \sim \omega^{-1}$ GeV, and $M \sim 10^5 \omega^{-1}$ (corresponding to the largest possible decay rate of leptoquark), we get the lower bound value for the leptoquark mass which is around $10^6$ GeV.  

\begin{figure}[t]
	\centering
	%\begin{minipage}{.3\textwidth}
	%	\centering
	%	\includegraphics[width=.9\linewidth]{decay1new}
	%	\subcaption{}\label{fig:1a}
	%	\caption{Process  $\rho^{\dagger}\rightarrow D_{L} + \bar{\nu}_{1R}$}
		%\label{figure1}
	%\end{minipage}%
	%\begin{minipage}{.3\textwidth}
	%	\centering
	%	\includegraphics[width=.9\linewidth]{decay2new}
	%	\subcaption{}\label{fig:1b}
	%	\caption{Process  $\rho\rightarrow d_{R} + \bar{\nu}_{1L}$}
		%\label{figure2}
	%\end{minipage}
	\begin{minipage}{.32\textwidth}
		\centering
		\includegraphics[width=.95\linewidth]{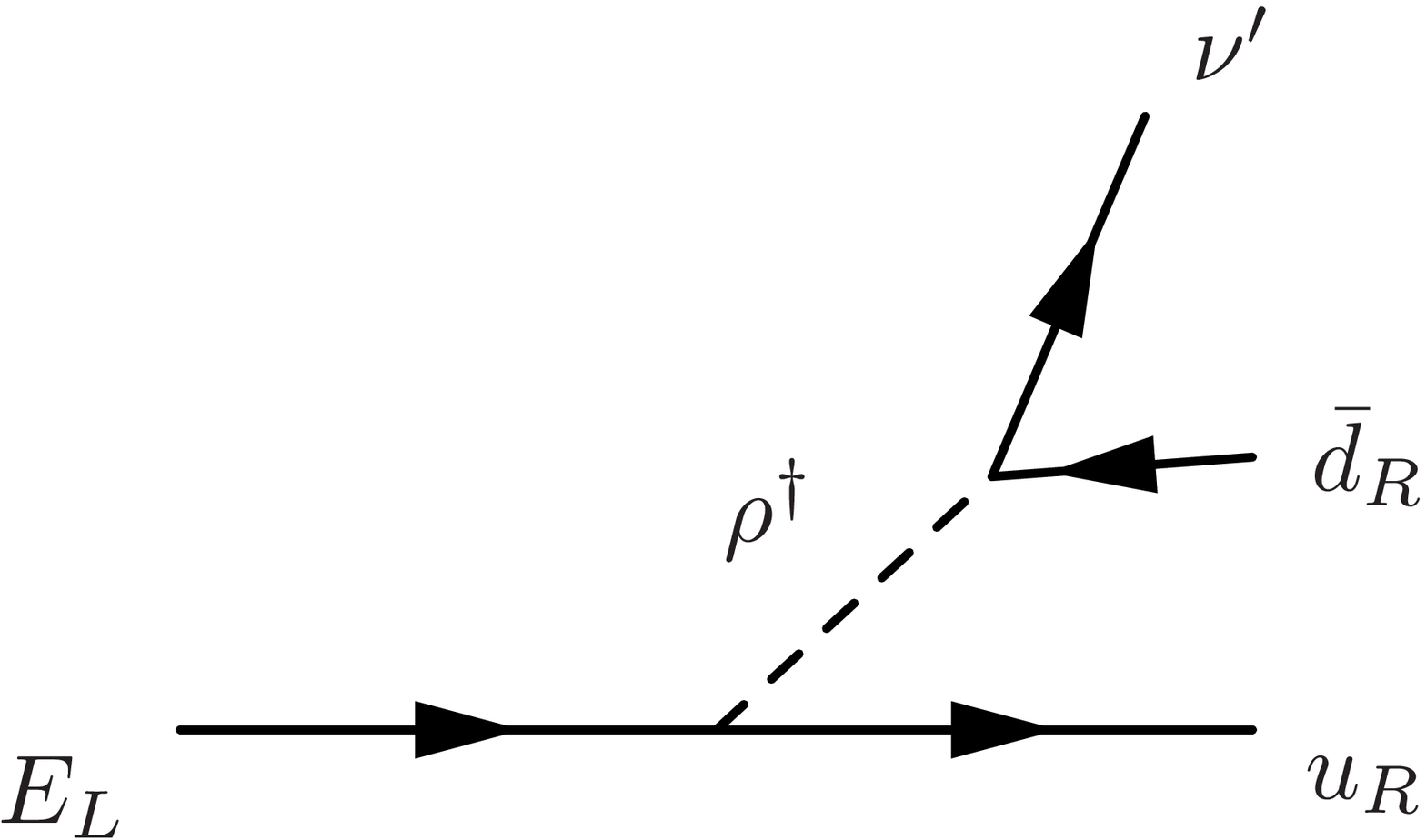}
		\subcaption{}\label{fig:1c}
	%	\caption{Process  $\rho^{\dagger}\rightarrow D_{L} + \bar{\nu}_{1R}$}
		%\label{figure1}
	\end{minipage}%
	%\\
	%\begin{minipage}{.3\textwidth}
	%	\centering
	%	\includegraphics[width=.9\linewidth]{decay4new}
	%	\subcaption{}\label{fig:1d}
	%	\caption{Process  $\rho\rightarrow d_{R} + \bar{\nu}_{1L}$}
		%\label{figure2}
	%\end{minipage}
	\begin{minipage}{.32\textwidth}
		\centering
		\includegraphics[width=.95\linewidth]{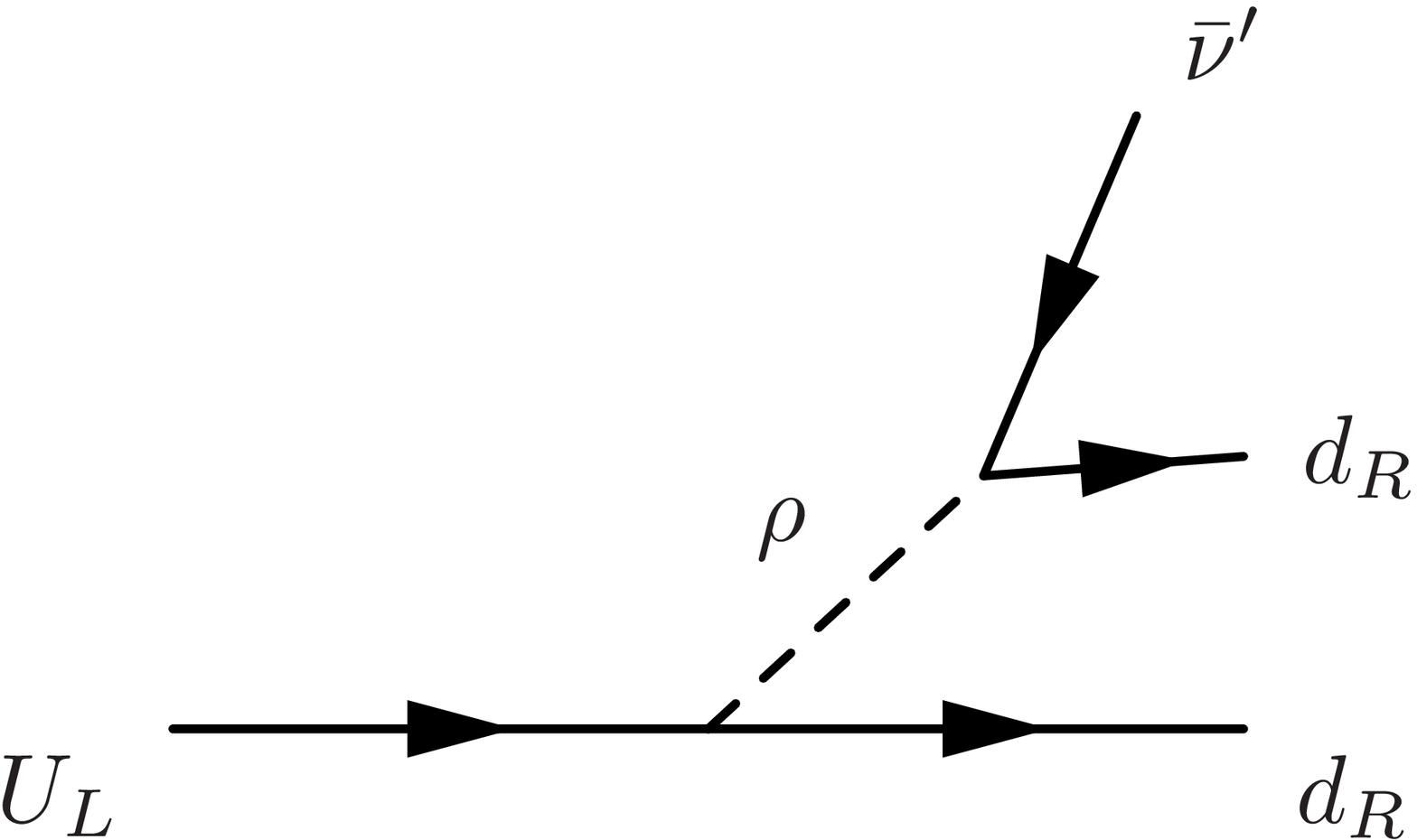}
		\subcaption{}\label{fig:1e}
	%	\caption{Process  $\rho^{\dagger}\rightarrow D_{L} + \bar{\nu}_{1R}$}
		%\label{figure1}
	\end{minipage}%
	\begin{minipage}{.32\textwidth}
		\centering
		\includegraphics[width=.95\linewidth]{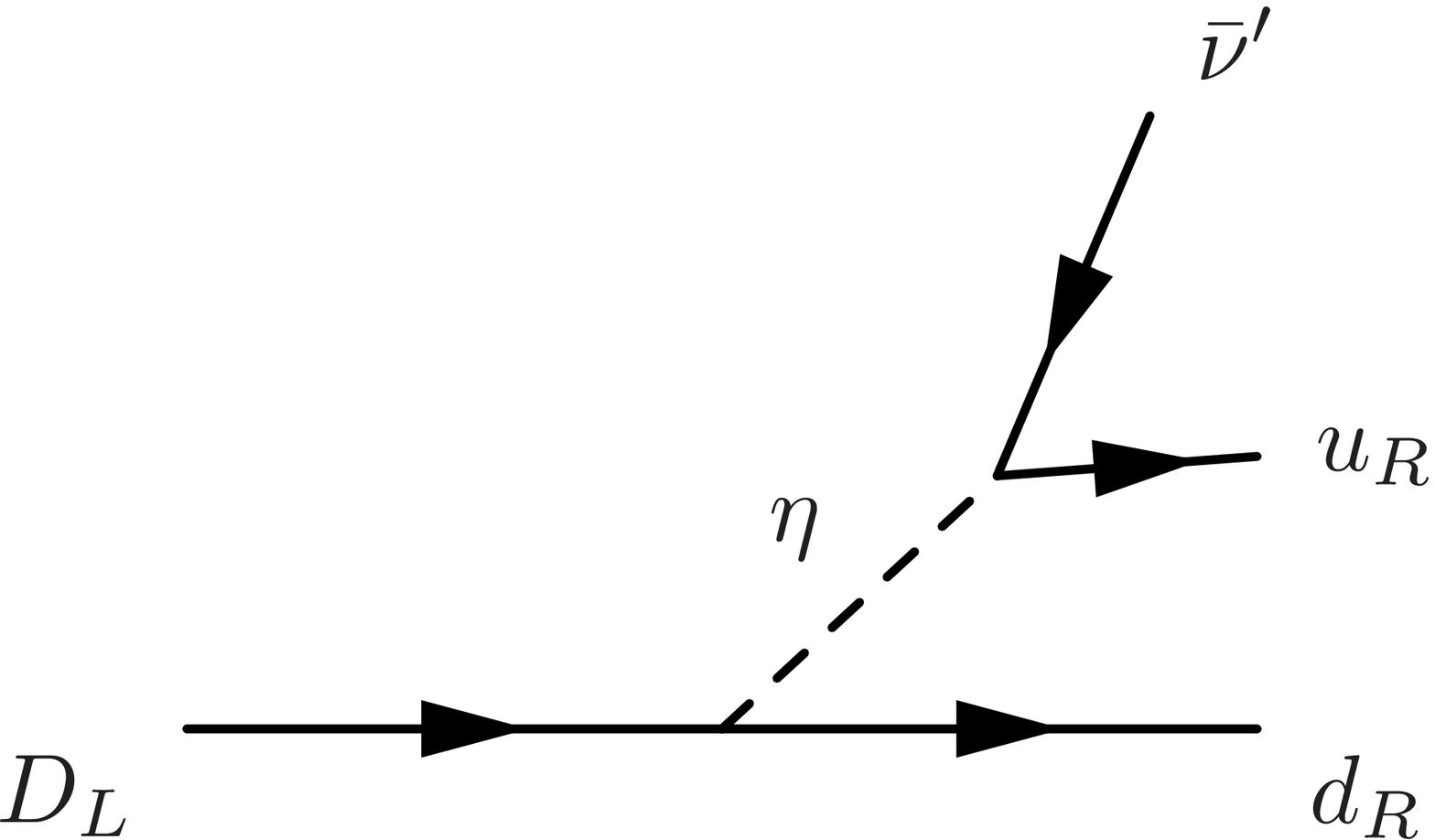}
		\subcaption{}\label{fig:1f}
	%	\caption{Process  $\rho\rightarrow d_{R} + \bar{\nu}_{1L}$}
		%\label{figure2}
	\end{minipage}
	\\
	\begin{minipage}{.32\textwidth}
		\centering
		\includegraphics[width=.95\linewidth]{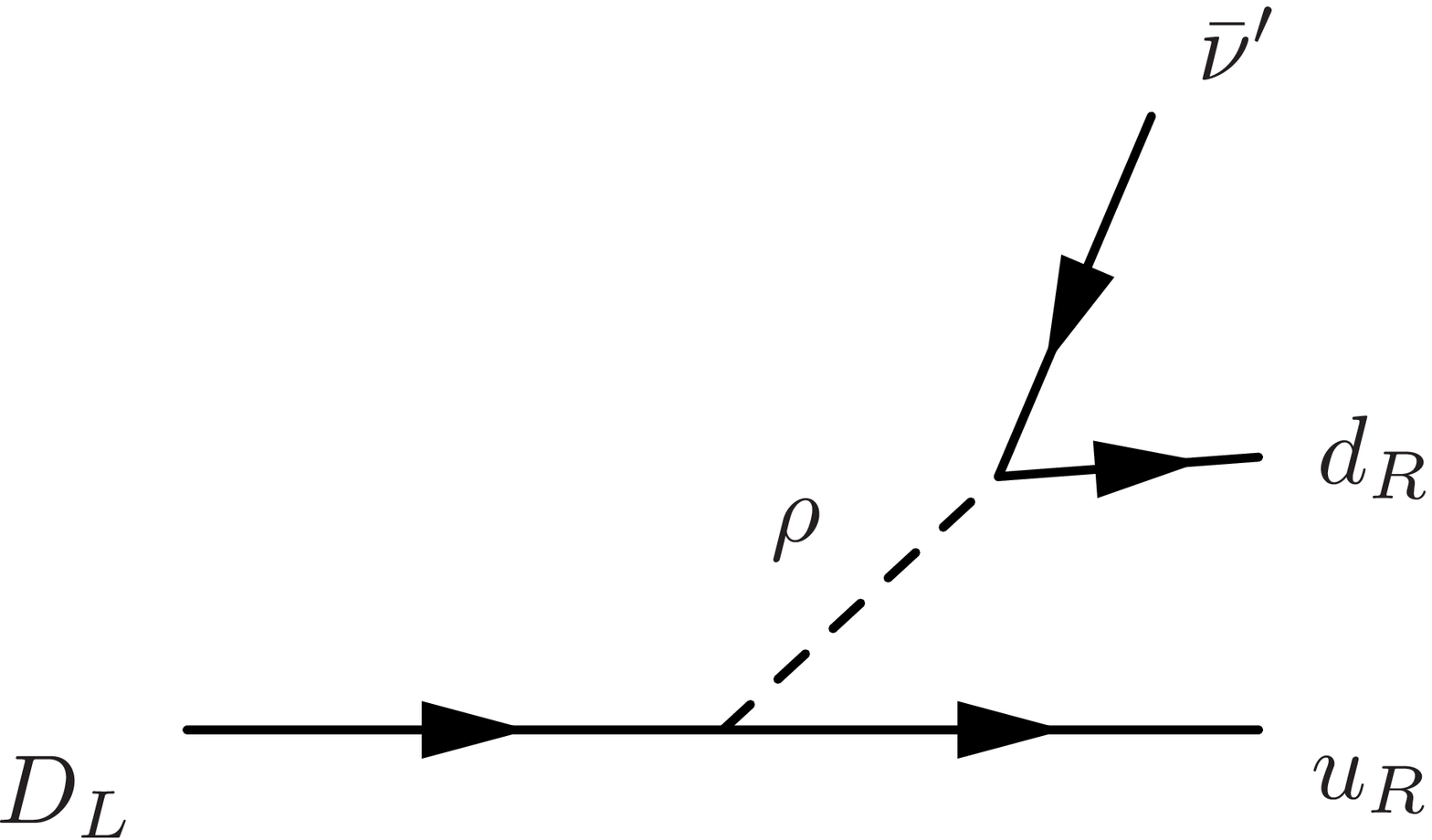}
		\subcaption{}\label{fig:1g}
	%	\caption{Process  $\rho^{\dagger}\rightarrow D_{L} + \bar{\nu}_{1R}$}
		%\label{figure1}
	\end{minipage}%
	%\\
	\begin{minipage}{.32\textwidth}
		\centering
		\includegraphics[width=.95\linewidth]{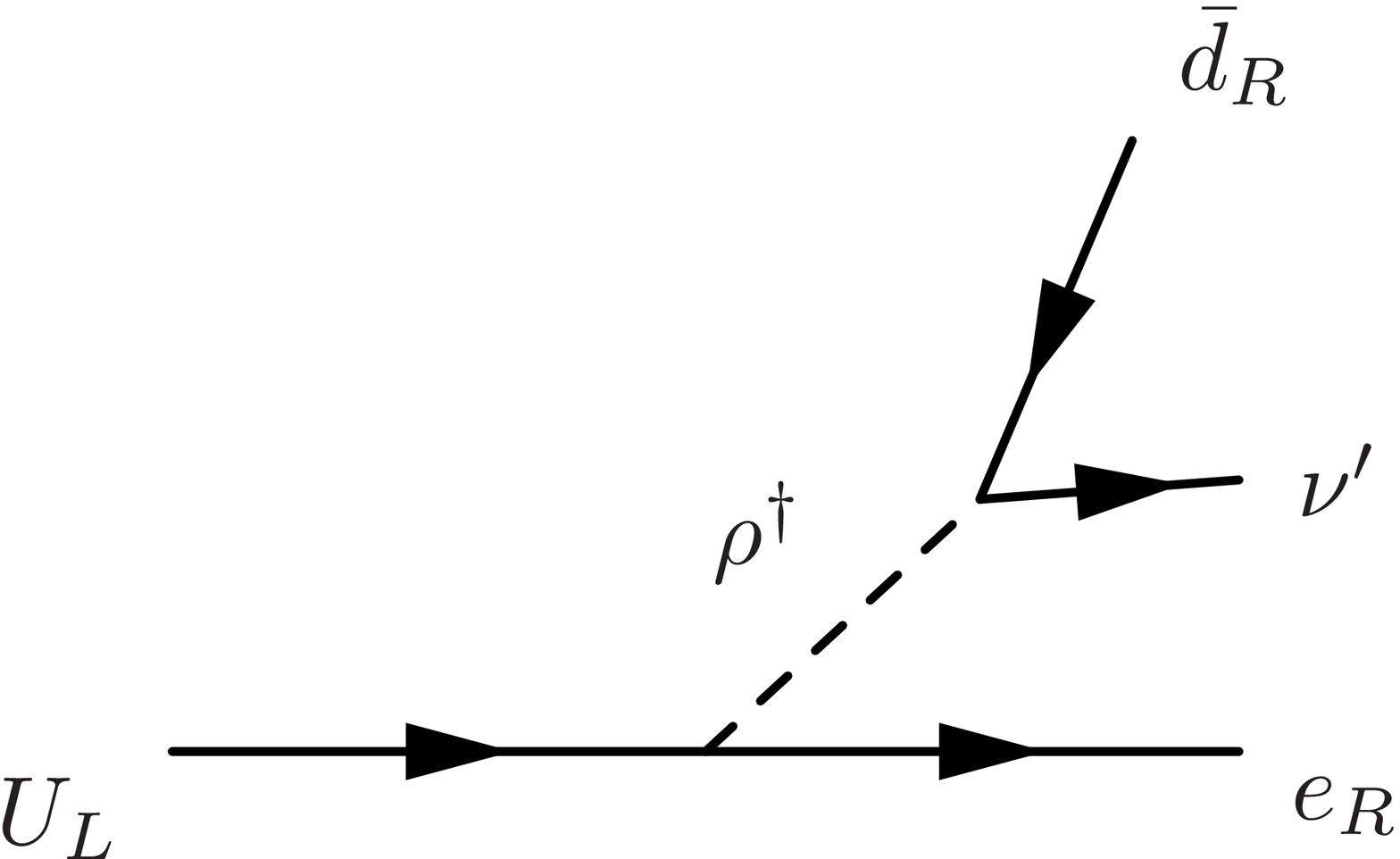}
		\subcaption{}\label{fig:1h}
	%	\caption{Process  $\rho\rightarrow d_{R} + \bar{\nu}_{1L}$}
		%\label{figure2}
	\end{minipage}
	%\begin{minipage}{.3\textwidth}
	%	\centering
	%	\includegraphics[width=.9\linewidth]{decay9new}
	%	\subcaption{}\label{fig:1i}
	%	\caption{Process  $\rho\rightarrow d_{R} + \bar{\nu}_{1L}$}
		%\label{figure2}
	%\end{minipage}
	\caption{The decay of the L-R dual fermions into the SM fermions via the leptoquarks}
\label{fig:fig}
\end{figure}

The leptoquarks are also needed in this model for facilitating the decay of massive L-R dual particles into  SM particles.  Due to the new LRS transformation, the same CKM matrix pattern should also occur in the L-R dual quark sector, thus any heavier generation of the L-R dual quark should decay into its  lighter generation via right charged gauge bosons.  In the lepton sector, any heavier generation of the L-R dual charged leptons should decay into its lighter generation and the lighter neutrinos.  Thus at  lower energy, we will end up with the lighter generation of L-R dual quarks, charged leptons and neutrinos.  The lightest generation of the $U$, $D$ quarks and the $E$ can decay into the SM fermions via the leptoquarks.  Figure \ref{fig:fig} shows the lowest order diagrams for the decay of L-R dual fermions.  

%The same CKM matrix pattern should also occur in the L-R dual quark sector due to the L-R dual symmetry, thus any heavier generation of the L-R dual quark should decay into its  lighter generation via right charged gauge bosons.  In the lepton sector, any heavier generation of the L-R dual charged leptons should decay into its lighter generation and the lighter neutrinos.  Thus at  lower energy, we will end up with the lighter generation of L-R dual quarks, charged leptons and neutrinos.  The lightest generation of the $U$, $D$ quarks and the $E$ can decay into the SM fermions via the leptoquarks.  Figure \ref{fig:fig} shows the lowest order diagrams for the decay of L-R dual fermions.  

At the lowest order, the decay rate of the L-R dual particles can be calculated easily.  Assuming the decayed products are massless and the leptoquarks $\rho$ and $\eta$ are very massive and of the same order, we get the decay rates of the L-R dual particles via leptoquark which are approximately given by,
\begin{equation}\label{decayratedual}
\Gamma \simeq \frac{M_{D}^5 |G_{ud}|^2 |G_{d\nu}|^2 }{M_{lq}^4} |v_R G_\nu M^{-1}|^2,
\end{equation} 
where $M_D$ is the mass of the (lightest generation) L-R dual particles.  The mass of the lightest quarks and charged leptons is in the order of $10^{-3}$ GeV, thus $M_D$ is in the order of $ 10^{-3}\omega^{-1}$ GeV.  Using the assumption for the neutrinos in the previous section, i.e. $M \sim \omega^{-1}10^5$, together with $v_R \sim \omega^{-1} v_L$, $G_\nu v_L \sim 1$  GeV, and $G_{ud}\sim G_{d\nu} \sim 1$, the lifetime of the L-R dual particles is approximately,
\begin{equation}\label{decayratedual2}
\tau = \Gamma^{-1} \simeq \omega^{5} M_{lq}^4 10^{25} \text{GeV}^{-1}\simeq \omega^5 M_{lq}^4 \text{s}  ,
\end{equation} 
where $M_{lq}$ above is in GeV unit.  If the leptoquarks are too massive, the decay of L-R dual particles are very slow and they should still remain in the universe.  For the case of $\omega = 10^{-6}$ and $M_{lq} = 10^6$ GeV, we have $\tau \simeq 10^{-6}$ s, which means the L-R dual particles have decayed away at the current age of the universe.  If the leptoquarks decayed away before the big bang nucleosynthesis (BBN), which is around several second after the big bang, then for the same $\omega$ the upper bound for $M_{lq}$ is around $10^7 \sim 10^8$ GeV.    

In many models that contain leptoquark, the proton can decay via a tree-level diagram. But in our model, this cannot happen since the leptoquarks always couple to fermion with massive L-R dual fermions. 
The existence of the L-R dual neutrinos with the same masses as the SM neutrinos clearly will add an additional relativistic degree of freedom during the BBN.  This can be avoided if the L-R dual sector is somehow colder than the SM sector.
In particular, the temperature ratio between L-R dual neutrino sector and the SM neutrino sector, $T'/T$, should be small.   If the first generation L-R dual particles (except neutrinos) energy density dominates the universe before decaying, then their decay into $\nu'$ and the SM particles will generate a large additional entropy to both sectors. Although $\nu'$ has the equal component of $\nu_L$ and $N_R$, their production depends on the strength of the weak interaction in  left and right sectors (i.e. depends on the ratio of $M_{W_L}$ to $M_{W_R}$).  Thus the ratio of $\nu_L$ to $N_R$ production is approximately proportional to the ratio of $\omega^{-4}$, which is very large.  Therefore the entropy generation to the SM sector will be much greater than the one to the L-R dual neutrino sector and $T'/T \ll 1$, thus the BBN constraint is avoided.

%\begin{figure}%[t]
%		\centering
%		\includegraphics[width=.55\linewidth]{2loop}
%	\caption{The two-loop $W_{L}-W_{R}$ mixing}
%\label{twoloop}
%\end{figure}

\section{Conclusion}
\label{sec5}

A new variant of L-R symmetry model has been proposed, where the L-R transformation is different than in the original LRS model, i.e. both the L-R chirality and the local quantum number are reversed.  The model also has an additional global quantum number $F$, which is unaffected with the new LRS transformation.    The particle contents are doubled, where for each SM particles we have its L-R duals.  The global quantum number $F$ will make the L-R dual particles different from the antiparticles of the SM particles. The model does not contain a bidoublet scalar, thus there is no mixing between left and right weak gauge bosons.  The existence of the global quantum number $F$ forbids the model from having any Majorana neutrinos.  All neutrinos in the model are Dirac neutrinos, but the seesaw mechanism and the usual SM neutrino oscillation can still occur.  Leptogenesis from the decay of Majorana neutrinos certainly is not available in this model, but baryon asymmetry can still be generated due to baryogenesis from the decay of  the leptoquarks.  Besides for baryogenesis, the leptoquarks are also required for the decay of L-R dual particles into SM particles.  The decay of  L-R dual particles will also produce a large entropy to the SM sector and gives a mechanism for avoiding the BBN constraint.

%Another interesting study is the possibility to investigate baryogenesis through scattering process as was suggested in 

%We found that the total of the baryon asymmetry is $B \sim (10^{-10}-10^{-6})$. 

%as well. We found that  
%Based on the discussion and calculation that have been performed, we can conclude several conclusions, namely
%\begin{enumerate}
%	\item Leptoquark particle can play a role as mediator for baryogenesis scenario in CP Mirror model. 
%	\item The formula of baryon asymmetry in CP Mirror model is given by
%	\begin{eqnarray}
%	B = 1,65\times 10^{-2}\ \epsilon \nonumber 
%	\end{eqnarray}
%	where 
%	\begin{eqnarray}
%	\epsilon = \frac{{\rm Im}(G_{d\nu}G_{du}^{*}G_{dd}G_{u\nu}^{*})}{3\pi}\left[\frac{\left\{1-\sigma^{2}\ln\left(1+1/\sigma^{2}\right)\right\}}{G_{d\nu}^{2}}-\frac{\left\{1-\sigma^{-2}\ln\left(1+\sigma^{2}\right)\right\}}{G_{u\nu}^{2}}\right] \nonumber
%	\end{eqnarray}
%	\item The number of baryon asymmetry is $B \sim (10^{-10}-10^{-6})$.
%	\item The lower bound of the mass of leptoquark particle is given by
%	\begin{eqnarray}
%	m_{lq} \gtrsim (10^{15}-10^{17})\  \text{GeV}. \nonumber
%	\end{eqnarray}
%\end{enumerate}

\begin{acknowledgments}
The work of A. S. Adam is supported by Hiroshima University research assistant fellowship. 

\end{acknowledgments}

%\appendix
%\include{appendix}

%\bibliography{my_reference}

\begin{thebibliography}{999}

\bibitem{patisalam2}
J. C. Pati and A. Salam, Phys. Rev. D {\bf 10}, 275 (1974)

\bibitem{mohapati}
R. N. Mohapatra and J. C. Pati, Phys. Rev. D {\bf 11}, 2558 (1975)

\bibitem{senjamoha}
G. Senjanovic and R. N. Mohapatra, Phys. Rev. D {\bf 12}, 1502 (1975)

\bibitem{senjanovic2}
G. Senjanovic, Nucl. Phys. B {\bf 153}, 334 (1979)

\bibitem{coutinho}
Y. A. Coutinho, J. A. M. Simoes and C. M. Porto, Eur. Phys. J. C {\bf 18}, 779 (2001)

\bibitem{simoes}
J. A. M. Simoes and J. A. Ponciano, Eur. Phys. J. C {\bf 30}, 007 (2003)

\bibitem{almeida}
F. M. L. de Almeida,  Y. A. Coutinho, J. A. M. Simoes, A. J. Ramalho, L. R. Pinto, S. Wulck and M.A. B. do Vale, Phys. Rev. D {\bf 81}, 053005 (2010)

\bibitem{seesaw}
M. Gell-Mann, P. Ramond and R. Slansky, in Supergravity, ed. by P. van Nieuwenhuizen
and D. Z. Freedman (North Holland, Amsterdam, 1979), p. 315; T. Yanagida, in Proc.
of the Workshop on the Unified Theory and Baryon Number in the Universe, ed. by
O. Sawada and A. Sugamoto (KEK report 79-18, 1979), p.95, Tsukuba, Japan; R.N.
Mohapatra and G. Senjanovic, Phys. Rev. Lett. 44 (1980) 912.


\bibitem{sirunyan}
A. M. Sirunyan \textit{et al.} (CMS Collaboration), Eur. Phys. J. C {\bf 79}, 421 (2019), arXiv:1809.10733 [hep-ex]; M. Aaboud \textit{et al.} (ATLAS Collaboration), Phys. Rev. D {\bf 99}, 072001 (2019),  arXiv:1811.08856 [hep-ex]

\bibitem{foot3}
R. Foot, H. Lew and R. Volkas, Phys. Lett. B {\bf 272}, 67 (1991)

\bibitem{volkas}
R. Foot and R. Volkas, Phys. Lett. B {\bf 645}, 75 (2007)

\bibitem{pdg}
 M. Tanabashi \textit{et al.} (Particle Data Group), Phys. Rev. D {\bf 98}, 030001 (2018). 

\bibitem{ronca}
M. Roncadelli and D. Wyler, Phys. Lett. B {\bf 133}, 325 (1983)

\bibitem{cabibo}
N. Cabibo, Phys. Rev. Lett. {\bf 10}, 531 (1963); M. Kobayashi and T. Maskawa, Prog. Theor. Phys. {\bf 49}, 652 (1973). 

\bibitem{weinberg3}
S. Weinberg, Cosmology. Oxford University Press, Oxford, UK (2008)

\bibitem{sakharov}
A. D. Sakharov, Pisma Zh. Eksp. Teor. Fiz. {\bf 5}, 32 (1967)

\bibitem{kolbturner}
E. W. Kolb and M. S. Turner, The Early Universe, Front. Phys. Vol. 69. Addison-Wesley (1990)

\bibitem{baryonnumber}
P. A. R. Ade \textit{et al.} (Planck Collaboration), Astron. Astrophys.{\bf 594}, A13(2016), arXiv:1502.01589  [astro-ph.CO].

%\bibitem{Nanowein}
%D. V. Nanopoulos and S. Weinberg, Phys. Rev. D {\bf 20}, 2484 (1979)



%\bibitem{riotto}
%A. Riotto, arXiv:hep-ph/9807454 (1998)

%\bibitem{bento}
%L. Bento and Z. Berezhiani, Phys. Rev. Lett. {\bf 87}, 231304 (2001)

\end{thebibliography}

\end{document}